\documentstyle[prl,aps,graphics]{revtex}
\begin{document}
\def\Universita{Universit\'a}
\def\Paris{Par\'{\i}s}
\def\Perez{P\'erez}
\def\Gunther{G\"unther}
\def\Schutzhold{Sch\"utzhold}
\def\Lofstedt{L\"{o}fstedt}
\def\Garcia{Garc\'\i{}a}
\def\ng{n_{\mathrm gas}}
\def\ngi{n_{\mathrm gas}^{\mathrm in}}
\def\ngo{n_{\mathrm gas}^{\mathrm out}}
\def\nl{n_{\mathrm liquid}}
\def\ni{n_{\mathrm in}}
\def\no{n_{\mathrm out}}
\def\nis{n_{\mathrm inside}}
\def\nos{n_{\mathrm outside}}
\def\Ni{ {\cal N}_{\mathrm in}}
\def\No{ {\cal N}_{\mathrm out}}
\def\max{\hbox{max}}
\def\min{\hbox{min}}
\def\in{{\mathrm in}}
\def\out{{\mathrm out}}
\def\sinc{\mathop{\hbox{sinc}}}
\def\half{{\textstyle{1\over2}}}
\def\quarter{{\textstyle{1\over4}}}
\def\omegai{\omega_\in}
\def\omegao{\omega_\out}
\def\observed{ {\mathrm observed}}
\def\kl{k_{\mathrm liquid}}
\def\kli{k_{\mathrm liquid}^{\mathrm in}}
\def\klo{k_{\mathrm liquid}^{\mathrm out}}
\def\kgi{k_{\mathrm gas}^{\mathrm in}}
\def\kgo{k_{\mathrm gas}^{\mathrm out}}
\def\vkgi{\vec k_{\mathrm gas}^{\; \mathrm in}}
\def\vkgo{\vec k_{\mathrm gas}^{\; \mathrm out}}
\def\vki{\vec k^{\; \mathrm in}}
\def\vko{\vec k^{\; \mathrm out}}
\def\ki{k^{\; \mathrm in}}
\def\ko{k^{\; \mathrm out}}

\twocolumn
\title{\bf Sonoluminescence:  
Bogolubov coefficients for the QED vacuum of a time-dependent
dielectric bubble.}
\author{Matt Visser$^{1,\P}$, S. Liberati$^{2,\dagger}$, 
F. Belgiorno$^{3,\ast}$, and D.W. Sciama$^{2,4,5,\S}$}
\address{$^1$ Physics Department, Washington University, 
Saint Louis MO 63130-4899, USA}
\address{$^{2}$ International School for Advanced Studies, Via Beirut 2-4, 
34014 Trieste, Italy}
\address{$^3$ \Universita\ degli Studi di Milano, Dipartimento di Fisica, 
Via Celoria 16, 20133 Milano, Italy}
\address{$^4$ International Center for Theoretical Physics,  
Strada Costiera 11, 34014 Trieste, Italy}
\address{$^5$ Physics Department, Oxford University, Oxford, England}
\date{8 May 1998; Revised 13 May 1999; \LaTeX-ed \today}
\maketitle

{\small

We extend Schwinger's ideas regarding sonoluminescence by explicitly
calculating the Bogolubov coefficients relating the QED vacuum states
associated with changes in a dielectric bubble. 
Sudden (non-adiabatic) changes in the refractive index lead to
an efficient production of real photons with a broadband spectrum, and 
a high-frequency cutoff that arises from the asymptotic behaviour of 
the dielectric constant. \\

 }


{\underline{\em Introduction:}} Sonoluminescence occurs when acoustic
energy induces the collapse of small bubbles, and the collapse of
these bubbles results in a brief intense flash of visible
light~\cite{Phys-Rep}. There are several competing mechanisms proposed
to explain this phenomenal concentration of kiloHertz acoustic energy
into PetaHertz electromagnetic energy.  An interesting mechanism,
originally proposed by Schwinger~\cite{Schwinger}, is based on changes
in the zero point fluctuations of the QED vacuum.  In his model,
Schwinger estimates the static Casimir energy of an expanded
dielectric bubble, compares it with the Casimir energy of a collapsed
dielectric bubble, and argues that this Casimir energy difference
(difference in zero point energies) would be converted into real
photons during collapse of the bubble. This model is often described
in terms of the dynamic Casimir effect (even if the calculation is
quasi-static).  Several authors have argued that this model is not
relevant to SL (cf.~\cite{qed} and references therein). Nevertheless,
we feel that the underlying idea of SL as a QED vacuum effect has been
prematurely discarded without sufficient analysis.

In this Letter we consider a variant of Schwinger's proposal which is
obtained by focusing our attention on changes in the refractive index
rather than on the bubble motion. We explicitly compute the Bogolubov
coefficients relating two vacuum states characterized by two different
values of the refractive index.  Non-trivial Bogolubov coefficients
imply the production of real photons.  The spectrum is qualitatively
compatible with those experimentally observed. Calculations are most
easily carried out for extremely large bubbles (large compared to the
cutoff wavelength), where the Bogolubov coefficients take on
particularly simple forms in terms of delta functions. The spectrum
and total energy emission are analytically calculable. For finite
bubbles, the delta functions are smeared by finite-volume effects and
the spectrum can be written down as an integral over a suitable sum of
spherical Bessel functions. This integral must be evaluated
numerically and can then be compared to both the large-volume
estimate, and to the experimental situation.  Even with a rather crude
(step function) model for the refractive index as a function of
frequency the resemblance between observed and predicted spectra is
quite reasonable. For technical details of the computation, and
additional discussion of the history of this proposal, see~\cite{qed}.


{\underline{\em Basic Features:}} One of the key aspects of photon
production by a space-dependent and time-dependent refractive index is
that for a change occurring on a timescale $\tau$, and a photon of
frequency $\omega$, then in the high frequency limit the amount of
photon production is exponentially suppressed by an amount
$\exp(-\omega\tau)$. The adiabatic approximation always implies such a
suppression~\cite{qed}.  The importance for SL is that the
experimental spectrum is {\em not\,} exponentially suppressed at least
out to the far ultraviolet.  Therefore any mechanism of
Casimir-induced photon production based on an adiabatic approximation
is destined to failure: Since the exponential suppression is not
visible out to $\omega \approx 10^{15} \hbox{ Hz}$, it follows that
{\em if\,} SL is to be attributed to photon production from a
time-dependent dielectric bubble ({\em i.e.}, the dynamical Casimir
effect), {\em then} the timescale for change in the dielectric bubble
must be of order a {\em femtosecond.}  Thus any Casimir--based model
has to take into account that {\em it is no longer the collapse from
$R_{\mathrm max}$ to $R_{\mathrm min}$ that is important}.

The SL flash is known to occur at or shortly after the point of
maximum compression. The light flash is emitted when the bubble is
at or near minimum radius $R_{\mathrm min} \approx 0.5\;\mu
\hbox{m} =500\;\hbox{nm}$. Note that to get an order femtosecond
change in refractive index over a distance of about $500\;{\rm nm}$,
the change in refractive index has to propagate at relativistic
speeds.  To achieve this, we must adjust basic aspects of the model:
We will move away from the original Schwinger suggestion and 
{\em we will postulate a rapid (order
femtosecond) change in refractive index of the gas bubble when it hits
the van der Waals hard core.}  The underlying idea is that there is
some physical process that gives rise to a sudden change of the
refractive index inside the bubble when it reaches maximum
compression. Given the fact that the timescale of such a change is
much shorter than that typical of the bubble collapse we shall
consider the bubble radius as fixed and equal to the minimal one
$R_{\mathrm min}$.  For the sake of simplicity we take, as Schwinger
did, only the electric part of QED, reducing the problem to a scalar
electrodynamics.  The equations of motion are
\begin{equation}
\epsilon {\partial^2\over\partial t^2} E-\nabla^{2} E=0.
\label{eqm}
\end{equation}
We shall consider two different asymptotic configurations for the gas
inside the bubble. An ``in'' configuration with refractive index
$\ni$, and an ``out'' configuration with a refractive index $\no$.
These two configurations will correspond to two different bases for
the quantization of the field. The two bases will be related by
Bogolubov coefficients in the usual way.  Once we determine these
coefficients we easily get the number of created particles per mode
and from this the spectrum.  Using the inner product
\begin{equation} 
(\phi_{1},\phi_{2}) =
i \int_{\Sigma_t} \epsilon(r,t) \; \phi_{1}^*
\stackrel{\leftrightarrow}{\partial}_{0}\phi_{2}\: d^{3}x,
\label{inpro}
\end{equation}
the Bogolubov coefficients are defined as
\begin{eqnarray}
\alpha_{ij}
&=&
({E_{i}^{\mathrm out}},{E_{j}^{\mathrm in}}), 
\qquad 
\beta_{ij}
=(
{E_{i}^{\mathrm out}}^*, {E_{j}^{\mathrm in}}).
\end{eqnarray}
We focus on the coefficient $\beta_{ij}$ because it is
related to the spectrum, number, and energy:
\begin{equation}
{dN \over d^3 \vec k_\out}
=\int|\beta(\vec k_\in ,\vec k_\out )|^{2}
\; d^3 \vec k_\in,
\end{equation}
\begin{equation}
N=\int {dN\over d\omega_\out} \; d\omega_\out ,
\end{equation}
\begin{equation}
E= \hbar \int {dN(\omega_\out )\over d\omega_\out }
\; \omega_\out  \; d\omega_\out .
\end{equation}
%


{\underline{\em Homogeneous dielectric:}} In the infinite volume limit
the eigen-modes are plane waves:
\begin{equation}
E(\vec x,t) = {1\over(2\pi)^{3/2}} \;
{\exp(i[\vec k\cdot \vec x - \omega t])\over
\sqrt{2 \; \omega}\; n}.  
\end{equation}
We now introduce a ``pseudo-time'' parameter by defining $ {\partial /
\partial \tau} = \epsilon(t) {\partial / \partial t}$, that is,
$\tau(t) = \int {dt/\epsilon(t)}$. Then
\begin{equation}
{\partial^2\over\partial\tau^2} E =
c^2 {\epsilon(\tau)}\nabla^2 E.
\label{eqm2}
\end{equation}
Now pick a convenient profile for the refractive index
\begin{eqnarray}
\label{E:profile}
\epsilon(\tau)&=& \half (\ni^2 + \no^2) +
\half(\no^2-\ni^2)\;\tanh(\tau/\tau_0).
\end{eqnarray}
Here $\tau_0$ represents the typical (pseudo-time) timescale for the
change of the refractive index.  After computation of the Bogolubov
coefficient we have to convert back to physical time.  Defining
$\langle n^2 \rangle=(\ni^2+\no^2)/2$ we get:
\begin{eqnarray}
&&
\left|\beta(\vki , \vko )\right|^2={V\over (2\pi)^3 } \;
\delta^3(\vki  + \vko  )
\nonumber\\
&&
\times
{\sinh^2\left(
{\textstyle \pi } 
{\textstyle |\ni^2 \omega_\in -\no^2 \omega_\out|} \; t_0
/ (2 \langle n^2 \rangle) 
\right)
\over
\sinh\left(
\pi \; {\textstyle \ni^2 } \;
\omega_\in t_0 / {\textstyle \langle n^2 \rangle} 
\right) \;
\sinh\left(
\pi \; {\textstyle \no^2} \;
\omega_\out t_0 /  {\textstyle \langle n^2 \rangle}
\right)
},
\label{bog2}
\end{eqnarray}
where $t_{0}$ is now the physical timescale of the change in
the refractive index.  For the particular temporal profile we have
chosen for analytic tractability this evaluates to $t_0 = \half \tau_0
\left( \ni^2 + \no^2 \right)$. The sudden approximation is valid
provided
\begin{equation}
\omega \ll \Omega_{\mathrm sudden} = {1\over 2\pi t_0} \;
{\ni^2+\no^2\over \no \; \max\{\ni,\no\}
}.
\end{equation}
Thus the frequency up to which the sudden approximation holds is not
just the reciprocal of the timescale of the change in the refractive
index: there is also a strong dependence on the initial and final
values of the refractive indices. This implies that we can relax, for
some ranges of values of $\ni$ and $\no$, our figure of $t_0\sim
O({\rm fs})$ by up to a few orders of magnitude.  Unfortunately the
precise shape of the spectrum is heavily dependent on all the
experimental parameters ($K,\ni,\no,R$). This discourages us from
making any sharp statement regarding the exact value of the physical
timescale required in order to fit the data.  In the region where the
sudden approximation holds the various $\sinh(x)$ functions in
equation (\ref{bog2}) can be replaced by their arguments $x$. Then
\begin{equation}
\left|\beta(\vki, \vko)\right|^2 \approx
{1\over 4}
{ (\ni - \no)^2 \over
\ni \; \no }\;
{V\over (2\pi)^3 } \; \delta^3(\vki + \vko ).
\label{E:sbsq}
\end{equation}
In the real physical situation $\ni$ is a function of $\omega_\in $
and $\no$ is a function of $\omega_\out $.  Schwinger's sharp
momentum-space cutoff for the refractive index is equivalent to the
choice
\begin{equation}
n(k) = n \; \Theta(K -k) + 1 \; \Theta(k-K).
\end{equation}
More complicated models for the cutoff are of course possible at the
cost of obscuring the analytic properties of the model.  Taking into
account the two photon polarizations
\begin{eqnarray}
&&|\beta(\vki,\vko )|^{2}\approx
{1 \over 2} \frac{\left(\no-\ni\right)^2}{\ni \no}
{V\over(2\pi)^3}\nonumber\\
&&
\qquad
\times \Theta(K -\ki ) \;  \Theta(K - \ko ) \;
\delta^3(\vki  + \vko ).
\label{E:largeb2}
\end{eqnarray}
The spectrum is
\begin{eqnarray}
&&{dN(\omega_\out )\over d\omega_\out }
\nonumber\\
&&\approx
{\no \over 2\; c}\;
{\left(\no-\ni\right)^2\over\no\,\ni}
{V\over(2\pi)^3}\; 4\pi \; (\ko)^2 \;
\Theta(K -\ko ).
\label{spc}
\end{eqnarray}
Thus at low frequencies, where the sudden approximation holds
strictly, the spectrum should show a polynomial behaviour (instead of
the linear one expected for a thermal distribution).
\footnote{This statement is independent of the explicit form of the 
profile for the change of the refractive index. Only very special
profiles (exponential and never-ending in time) provide an exactly
thermal spectrum~\cite{BLH}.}
The number of emitted photons is
\begin{eqnarray}
N
&\approx&
{1\over 9\pi} \;
{\left(\no-\ni\right)^2\over\no\ni} \;
(R K)^3.
\end{eqnarray}  
The total emitted energy is approximately
\begin{eqnarray}
E&\approx&
{1\over16\pi^2} \;
{\left(\no-\ni\right)^2\over\ni\no^2} \;
\hbar\; c\; K \; {V K^3} =
\frac{3}{4}\; N\; \hbar \omega_{\mathrm max}.
\label{E:energy}
\end{eqnarray}
So the average energy per emitted photon is approximately $
\langle E \rangle = {3\over 4} \hbar  \; \omega_{\mathrm
max}\sim 3 \; {\rm eV}$. To extract some numerical estimates, recall
that in our new variant of Schwinger's model we have $R_{\mathrm
light-emitting-region} \approx R_{\mathrm min} \approx 500\;
\hbox{nm} $. We take $K_\observed = K \nl \approx 2\pi/(200\; {\mathrm nm})$
so that $K_\observed R\approx 5 \pi \approx 15$. To get about one
million photons we now need, for instance, $\ni\approx 1$ and
$\no\approx 12$, or $\ni\approx 2\times 10^4$ and $\no\approx 1$, or
even $\no \approx 25$ and $\ni\approx 71$, though many other
possibilities could be envisaged.  Note that the estimated values of
$\ngo$ and $\ngi$ are extremely sensitive to the precise choice of
cutoff, and the size of the light emitting region.

\begin{figure}[htb]
\vbox{
\hfil
\scalebox{0.66}{\includegraphics{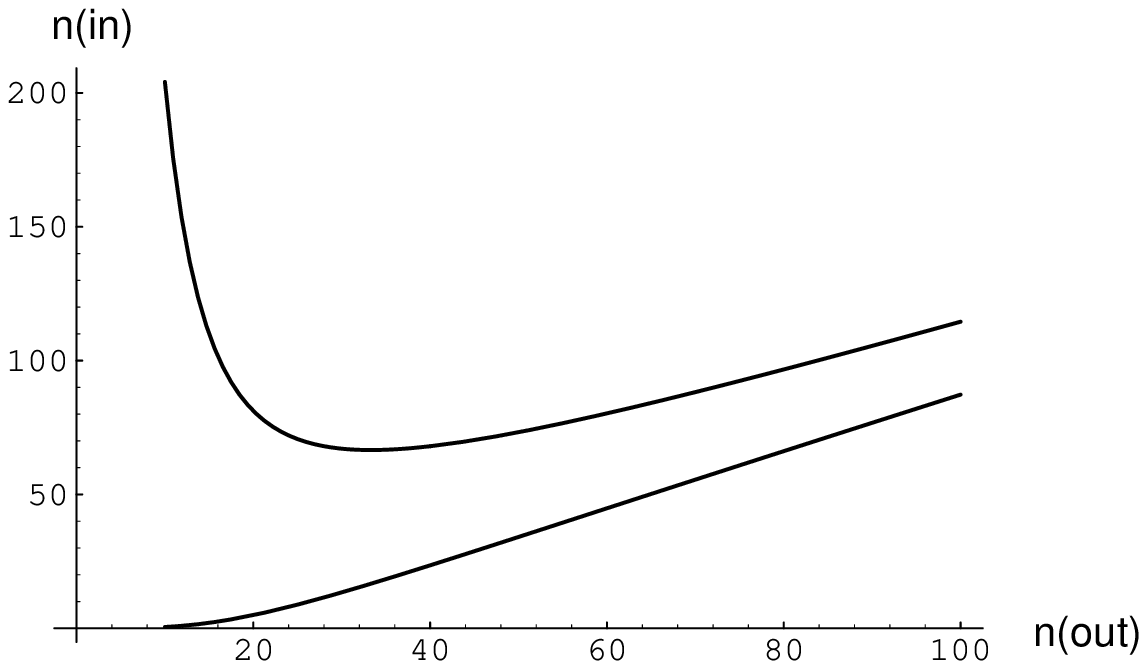}}
\hfil
}
\caption{%
The initial refractive index $\ni$ plotted as a function of $\no$ when
one million photons are emitted in the sudden approximation.}
\label{F:photons-1} 
\end{figure}

More systematically, using $K_\observed R \approx 15$ we get
\begin{equation}
N = {119\over\nl^3}\; (\no-\ni)^2 \; {\no^2 \over \ni}.
\end{equation}
Solving for $\ni$ as a function of $\no$ and $N$, and taking $N=10^6$,
the result is plotted in figure (\ref{F:photons-1}). For any specified
value of $\no$ there are exactly two values of $\ni$ that lead to one
million emitted photons.

{\underline{\em Finite volume effects:}} In finite volume the
eigenmodes are combinations of Bessel functions and Spherical
Harmonics, subject to
\begin{equation}
\epsilon=\left \{ 
\begin{array}{llll}
\epsilon_{1} 
& = & \epsilon_{\mathrm bubble-contents} & \mbox{if $r< R$},\\
\epsilon_{2} 
& = & \epsilon_{\mathrm medium} & \mbox{if $r >R$},
\end{array}
\right.
\end{equation}
and satisfying appropriate junction conditions at $r=R$.  We limit
ourself to just quoting the key result \cite{qed}.  Introducing
$\Delta n\equiv \ngi-\ngo$, 
\begin{eqnarray}
&&{dN\over d\omega_\out} =
\quarter R^2 (\Delta n)^2 \; \sum_{l=1}^\infty (2l+1) \;
\nonumber\\
&& \qquad 
\times
\int d\omega_\in \left\{
{
\ngo\omega_\out^2 
+
\ngi\omega_\in^2
\over
\omega_\out+\omega_\in
}\right\}^2
 \left| A_{\nu}^\in \right|^{2}
 \left| A_{\nu}^\out \right|^{2}
\nonumber\\
&&
\qquad
\times
 \left[  
        {W[J_{\nu}(\ngo\,\omega_\out r/c),
           J_{\nu}(\ngi\,\omega_\in r/c)]_{R}
        \over
        (\ngo\,\omega_\out)^2-(\ngi\,\omega_\in)^2}
  \right]^2.
\label{E:b2}
\end{eqnarray}
Where $W(f,g)$ is the Wronskian and the $A_\nu^\in$ are calculable
coefficients depending on the frequency, refractive index of the
bubble, and that of the ambient medium.  The above is a general result
applicable to any dielectric sphere that undergoes sudden change in
refractive index. This expression is far too complex to allow a
practical analytical resolution of the general case.  For the specific
case of sonoluminescence, we have developed suitable numerical
approximations.  In the infinite volume limit there were two
continuous branches of values for $\ngi$ and $\ngo$ that led to
approximately one million emitted photons. If we now place the same
values of refractive index into the spectrum obtainable from the
Bogolubov coefficients derived above, numerical integration again
yields approximately one million photons. The total number of photons
is changed by at worst a few percent, while the average photon energy
($3/4$ times the cutoff energy) is almost unaffected. (Some specific
sample values are reported in Table I.) The basic result is this: as
expected, finite volume effects do not greatly modify the results
estimated by using the infinite volume limit. Note that $\hbar
\omega_{\mathrm max}$ is approximately $4$ eV, so that average photon
energy in this crude model is about $3$ eV.

\begin{center} 
\bigskip
\begin{tabular}{|c|c|c|c|}
\hline
$\;\ngi\;$ & $\;\ngo\;$  & Number of photons &
$\langle E \rangle/\hbar \omega_{\mathrm max}$ \\
\hline
\hline
$2\times10^4$ & $1$  & $1.06\times 10^6$ & $0.803$ \\
\hline
$71  $   & $25$ & $1.00\times 10^6$ & $0.750$ \\
\hline
$68  $   & $34$ & $1.06\times 10^6$ & $0.751$ \\
\hline
$9$   & $25$ & $0.955\times 10^6$ & $0.750$ \\
\hline
$1$      & $12$ & $0.98\times 10^6$ & $0.765$ \\
\hline
\end{tabular}
\medskip
\center{Table I: Some typical cases.}
\bigskip
\label{T:table} 
\end{center}
%
In addition, for the specific case $\ngi=2\times10^4$, $\ngo=1$, we
have calculated and plotted the form of the spectrum. We find that the
major result of including finite volume effects is to smear out the
otherwise sharp cutoff coming from Schwinger's step-function model for
the refractive index. Other choices of refractive index lead to
qualitatively similar spectra.  These results are in reasonable
agreement with experimental data.
%
\begin{figure}[htb]
\vbox{
\hfil
\scalebox{0.375}{\rotatebox{270}{\includegraphics{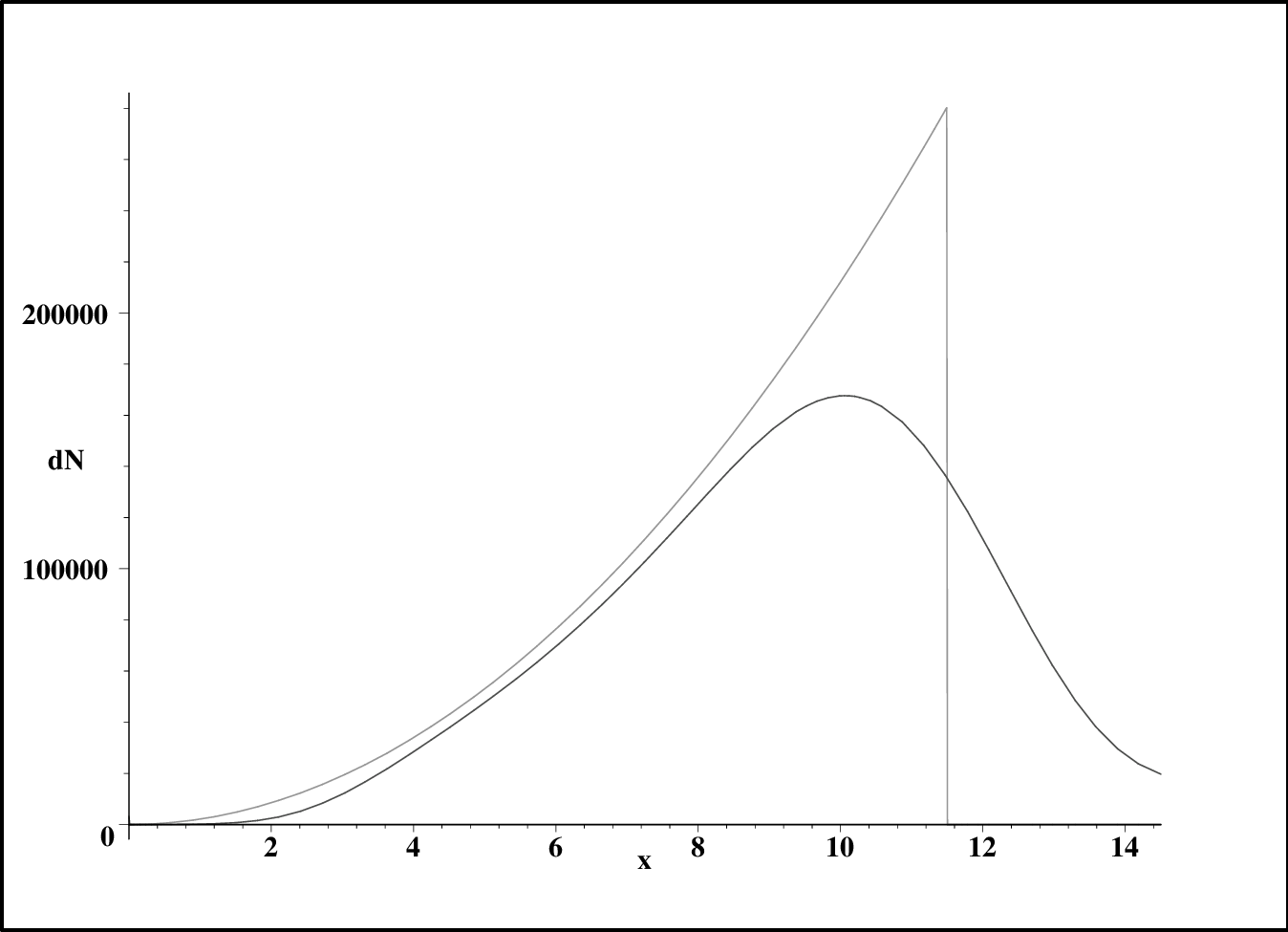}}}
\hfil
}
\bigskip
\caption{%
Spectrum $dN/dx$ obtained by integrating the approximated Bogolubov
coefficient.
For $\no=1$ and $R=500$nm the relation between the non-dimensional
quantity $x$ and the frequency $\nu$ is $x\sim \nu \cdot 10.5 \cdot
10^{-15}$s. So $x\approx 11.5$ corresponds to $\nu \approx 1.1$ PHz.
The curve with the sharp cutoff is the infinite volume
approximation. Finite volume effects tend to smear out the sharp
discontinuity, but do not greatly affect the total number of photons
emitted.
} 
\label{F:spectrum} 
\end{figure}


{\underline{\em Conclusions:}} We suggest that the key to the SL
phenomenon is not the details of the bubble collapse, but rather the
way in which the refractive index changes as a function of space and
time. Sudden changes in the refractive index will lead to efficient
conversion of zero point fluctuations into real photons.  Fitting the
details of the observed SL spectrum then becomes an issue of building
a robust model for such sudden changes of the refractive index of the
entrained gases as functions of frequency, density, and composition.

A viable conjecture is that the refractive index of the trapped gas
undergoes major changes near the moment of maximum compression, when
the molecules in the gas bounce off the van der Waals' hard
core. Femtosecond timescales for changes in the refractive index have
already been envisaged in the literature~\cite{Yablonovitch}, but we
feel that the present model should encourage a more detailed
experimental investigation regarding the dynamics of the refractive
index, both that of the entrained gas as well of the water surrounding
the bubble.

We stress that {\em any mechanism} that provides sub-picosecond
timescales for the change of the refractive index would imply an
important contribution of Casimir photons in sonoluminescence. It is
not inconceivable that for some such mechanisms ({\em e.g.}, sudden
ionization) and for some regions of the parameter space the dynamical
Casimir effect could play an important role in addition to or in
opposition to other proposed explanations of photon emission ({\em
e.g.}, Bremsstrahlung or shock waves).
 
Generic features for testing our proposal are~\cite{qed}, non-thermal
behaviour of the spectrum at low frequencies, absence of hard UV
photons (no dissociation of the water molecules), emission of photons
bounded in angular momentum ($l\leq KR$). Detection of
correlations~\cite{2gamma} in the emitted photons has also been
identified as a possibly efficient tool for discriminating between
vacuum-effect-based models of SL and thermal light emission models.

We argue, both here and elsewhere~\cite{qed,2gamma}, that Casimir-like
mechanisms for SL are viable, that they make both qualitative and
quantitative predictions, and that they are now sufficiently well
defined to be experimentally falsifiable.


\end{document}